\documentstyle[epsfig,longtable]{aipproc}

\begin{document}
\title{Isospin Breaking in the Extraction of Isovector and Isoscalar
Spectral Functions from $e^+e^-\rightarrow hadrons$}

\author{K. Maltman$^*$ and C.E. Wolfe$^\dagger$}
\address{$^*$Dept. Mathematics and Statistics, York Univ.,
4700 Keele St., Toronto, ON Canada, and CSSM, Univ. of Adelaide, 
Adelaide, SA Australia\thanks{Supported by the Natural Sciences
and Research Engineering Council of Canada} \\
$^{\dagger}$Nuclear Theory Center, Indiana University, Bloomington,
IN, USA}

%\lefthead{LEFT head}
%\righthead{RIGHT head}
\maketitle

\begin{abstract}
A finite energy sum rule (FESR) analysis of the isospin-breaking
vector current correlator 
$\langle 0\vert T\left( V^3_\mu V^8_\nu \right)\vert 0\rangle$
is used to determine the isospin-breaking electromagnetic (EM)
decay constants of the low-lying vector mesons.  These results
are used to evaluate the corrections required to extract
the flavor diagonal $33$ and $88$ resonance contributions from
the full resonance EM contributions to the EM spectral function.
A large ($\sim 15\%$) correction is found in the case of the
$\omega$ contribution to the isoscalar spectral function. 
The implications of these results for sum rules
based on the isovector-isoscalar spectral difference are considered.
\end{abstract}

\section*{Introduction}
If we define the flavor $ab$ ($a,b=3,8$) vector current correlators
via
\begin{equation}
i \int d^4x \, e^{iqx} 
\langle 0|T(V^a_{\mu}(x) V^b_{\nu}(0)^\dagger)|0\rangle \, 
\equiv (-g_{\mu\nu} q^2 + q_{\mu} q_{\nu}) \, \Pi^{ab}(q^2)\ ,
\label{correlatordefn}\end{equation}
the corresponding EM current-current correlator is given by
\begin{equation}
\Pi^{EM} = \Pi^{33}+{\frac{2}{\sqrt{3}}}\Pi^{38}
+{\frac{1}{3}}\Pi^{88}\ ,
\end{equation}
which reduces to $\Pi^{33}+\Pi^{88}/3$ in the isospin limit.
$\Pi^{EM}$ is related to the hadronic electroproduction
cross-section by
\begin{equation}
\sigma (e^+e^-\rightarrow hadrons)={\frac{16\pi^2\alpha^2}{s}}\, {\rm Im}
\Pi^{EM}(s)\ .
\nonumber
\end{equation}
In the isospin limit, the $33$ and $88$ contributions 
associated with $n\pi$
states can be classified by $G$-parity, making an experimental
determination of the separate isovector ($33$) and 
isoscalar ($88$) components
of the EM spectral function, $\rho^{EM}(s)$, possible.
Isospin breaking (IB), however, complicates this separation because
of the presence of a non-zero $38$ spectral component.
One contribution to the $38$ component, associated with the
process $e^+e^-\rightarrow\omega\rightarrow\pi^+\pi^-$,
is conventionally removed by hand in determining the 
$33$ spectral function, but other $38$ contributions also
exist which are not so subtracted.  If we define, in obvious
notation, the isovector $(3)$ and isoscalar $(8)$ components of the
EM decay constant of vector meson, $V$, via
\begin{equation}
F_V^{EM}=F_V^{3}+{\frac{1}{\sqrt{3}}}F_V^8
\label{ffdefn}
\end{equation}
then, for example, the intermediate $\rho$ contribution to $\rho^{EM}(s)$
contains not only a $33$ component,
proportional to $[F^3_\rho]^2$, but also, to leading
order in IB, a $38$ component proportional to $F_\rho^{3}F_\rho^{8}$.
The $\omega$ contribution to $\rho^{EM}(s)$ similarly contains a
$38$ component proportional to $F_\omega^{3}F_\omega^{8}$.

The presence of these $38$ ``contaminations'' in the conventionally
extracted $33$ and $88$ spectral functions creates potential problems
for 
\begin{itemize}
\item CVC tests
\item the determination of $m_s$ using Narison's $33-88$ ``tau-decay-like''
sum rule\cite{narisonms}
\item The inverse weighted chiral $33-88$ sum rule for the
$6^{th}$ order ChPT LEC, $Q_V$, which governs flavor/isospin breaking
in vector current correlators\cite{gkqvsr}.
\end{itemize}
In the latter two cases, the IB effect can be particularly important
(1) because the cancellation between the nominal
isovector and isoscalar integrals is typically rather strong
(to the $\sim 10\%$ level) and, (2) because 
the $38$ component of the $\rho$ and $\omega$ EM decay constants
induced by $\rho$-$\omega$ mixing has the opposite sign
for the $\rho$ and $\omega$, the $38$ corrections
add constructively when one forms the nominal isovector-isoscalar
difference.

It is easy to see that the relative importance
of this effect should be much larger 
for the isovector ``contamination''
of the $\omega$ decay constant than for the isoscalar ``contamination''
of the $\rho$ decay constant.  Indeed,
taking $SU(3)_F$ for the vector current
vacuum-to-vector-meson matrix elements, and assuming ideal mixing 
of the vector meson sector,
$F^{(0)}_\rho \simeq 3F^{(0)}_\omega$,
where the superscript $(0)$ indicates the isospin-unmixed states.
With $\epsilon$ the $\rho$-$\omega$ mixing angle, one then has
\begin{eqnarray}
F^{EM}_\rho&=&F^{(0)}_\rho -\epsilon F^{(0)}_\omega 
\simeq F^{(0)}_\rho\left( 1-\frac{\epsilon}{3}\right) \nonumber \\
F^{EM}_\omega&=&F^{(0)}_\omega +\epsilon F^{(0)}_\rho 
\simeq F^{(0)}_\omega\left( 1+3\epsilon \right) \ .
\label{rhoomegaexpectations}
\end{eqnarray}
The fractional contribution of the ``wrong-isospin'' current
to the EM decay constant of the $\omega$ should thus be $9$ times
that of the corresponding IB contribution to $F^{EM}_\rho$,
in the limit that the effect is dominated by $\rho$-$\omega$
mixing.

In this paper, we employ a FESR analysis of the IB vector
current correlator, $\Pi^{38}$, in order to extract
the product, $F_V^3F_V^8$, 
of IB and isospin-conserving (IC) decay constants for
the low-lying vector mesons.  This allows us to 
determine the $38$ contamination of the resonance contributions to
the nominally extracted $33$ and $88$ spectral
functions, and hence to make the corrections required to
extract the actual versions thereof from data.

\section*{FESR's and the Determination of the IB Vector Meson Decay Constants}
The analyticity structure of $\Pi^{38}$ is such that,
for any function $w(s)$ analytic
in the region of the contour, one has,
using Cauchy's theorem, 
the FESR relation
\begin{equation}
\int_{s_{th}}^{s_0}\, ds\, w(s){\rho^{38} (s)}={\frac{-1}{2\pi i}}
\oint_{\vert s\vert =s_0}\, ds\, w(s)\Pi^{38} (s) \ ,
\nonumber\end{equation}
where $\rho^{38}(s)={\frac{1}{\pi}}{\rm Im}\, \Pi^{38}(s)$.
Choosing $s_0$ large enough that $\Pi^{38}$ can be 
represented by the OPE on the RHS, but small enough that $\rho^{38}$
on the LHS is still dominated by the known isovector and
isoscalar vector mesons, one obtains a relation
between the products $f_V=F^3_VF^8_V$ which govern the sizes
of resonance contributions to $\rho^{38}$ and the parameters
($\alpha_s$ and vacuum condensates) entering the OPE.  Since the OPE
is well known this allows us to constrain the resonance parameters.

In order to make the analysis suggested above reliable, we will employ
a form of FESR tested in the isovector vector channel and shown to
be very accurately satisfied there\cite{kmfesr}.  These 
``pinch-weighted'' FESR's are those corresponding 
to weights satisfying $w(s_0)=0$.  In the isovector vector channel,
where they can be tested, they are known to be very accurately
satisfied down to rather low scales ($\sim 2\ {\rm GeV}^2$, even
when more general FESR's, which do not suppress contributions from
the region near the timelike real axis, are poorly satisfied\cite{kmfesr}).
It is convenient to work with two families of pinched weights,
$w_s(s)=[1-s/s_0][1+As/s_0]$ and $w_d(s)=[1-s/s_0]^2[1+As/s_0]$,
where in each case $A$ is a free parameter, used to vary the weight
profile.  It is worth noting that, using the FESR's resulting
from these two weight families to fit the decay constants
of the first three $\rho$ resonances, one obtains a determination
of the $\rho (770)$ decay constant in terms of OPE parameters which
is accurate to within experimental errors\cite{kmfesr,kma0}.

On the hadronic side of our FESR's we assume a sum of Breit-Wigner
resonance contributions for the $\rho$, $\omega$, and $\phi$
with PDG values of the masses and widths. In
the $\rho^\prime$-$\omega^\prime$ region, a
single effective resonance contribution
with the average values of the masses and widths
is employed since the individual $\rho^\prime$ and $\omega^\prime$
masses and widths are very similar,
making it impossible for our sum rule to discriminate
between them.
The necessity of including a
possible $\phi$ term has been discussed in Ref.~\cite{kmold38}, while
the importance of including the resonance widths has been stressed in
Ref.~\cite{ijl}).  
The only
remaining unknowns in the spectral ansatz are 
the products
$f_V=F^3_VF^8_V$, which determine the heights of the resonance
peaks.

The $D=0,2,4,6$ terms in the OPE representation of $\Pi^{38}$ are given by
\begin{eqnarray}
&&\left[ \Pi^{38}_{1\gamma E}\right]_{D=0}
= -{\frac{\alpha} {16\pi^3}}{\frac{1}{4\sqrt{3}}}ln(Q^2)
\nonumber \\
&&\left[ \Pi^{38}(Q^2)\right ]_{D=2}=
{\frac{3}{2\pi^2 Q^2}}{\frac{1}{4\sqrt{3}}}
\left[(m_d^2-m_u^2)(Q^2)\right]
\left [1+{\frac{8}{3}} a(Q^2)\right. \nonumber \\
&&\qquad\qquad\qquad \left. +\left (
{17981\over 432}+{62\over 27} \zeta(3)
-{1045\over 54} \zeta(5)\right ) a^2(Q^2)\right ]
\nonumber \\
&&\left[\Pi^{38}(Q^2)\right]_{D=4} =
{\frac{2
\left(\langle m_u\bar{u}u\rangle - \langle m_d\bar{d}d\rangle\right)}
{4\sqrt{3}Q^4}}
\left[ 1+{1\over 3}a(Q^2)+{11\over 2}a^2(Q^2)\right] \nonumber \\
&&\left[ \Pi^{38}(Q^2)\right]_{D=6} =
{\frac{112\pi}{81\sqrt{3}Q^6}}
\rho_{red}\gamma (\rho\alpha_s\langle\bar{q}q\rangle^2)\ ,
\label{OPEforms}
\end{eqnarray}
where $\alpha$ is the usual EM coupling, $a(Q^2) = \alpha_s(Q^2)/\pi$, 
$\zeta (n)$ is the Riemann zeta function, 
$\gamma = [<\bar{d}d>/<\bar{u}u>]-1$,
and $\rho_{red}$ is the ratio of the violation of the vacuum
saturation approximation (VSA) estimate for the $D=6$ condensate combination
in the $38$ channel to that in the flavor-diagonal $33$ vector
channel (see Ref.~\cite{narisonrho} for the latter
value).  Numerically, the $D=4$ term is the largest, owing to the
smallness of $\alpha$ and $m_{u,d}$.  It can be rewritten in
terms of the IB mass ratio $r\equiv (m_d-m_u)/(m_d+m_u)$
(for which we use the value from Leutwyler's ChPT
analysis\cite{leutwylerqm}) and $<(m_u+m_d)\bar{u}u>\simeq -f_\pi^2 m_\pi^2$
(where we have used the GMOR relation).  We employ the 4-loop 
versions of the running mass and coupling.

Lack of knowledge of
$\rho_{red}$ would normally limit the accuracy of the extraction of
the $f_V$.  Using both the $w_s$ and $w_d$ families of FESR's,
however, it turns out to be possible to determine $\rho_{red}$
self-consistently.  In Figure 1 we illustrate the dependence
of $f_\rho$ on $\rho_{red}$ for both the $w_s$ and $w_d$ weight families.
Obviously only one value of $\rho_{red}$ provides
a consistent determination of $f_\rho$.  Checking
analogous figures for the other $f_V$, one finds that the
values of $\rho_{red}$ for which they become consistent
agree with that in Figure 1 to better than $1\%$.  This
clearly demonstrates that this ``self-consistency'' determination
of $\rho_{red}$ is physically meaningful. 
It is interesting to note that the degree of VSA violation is
very similar for the flavor-diagonal isovector and IB flavor $38$
correlators.  Further details of the calculation, including
OPE input values, uncertainties, and a description of the
contour-improved implementation of the OPE integrals~\cite{cipt}, may
be found in Ref.~\cite{kmcewv3v8}.

\begin{figure}[b!] % fig 1
\centerline{\epsfig{file=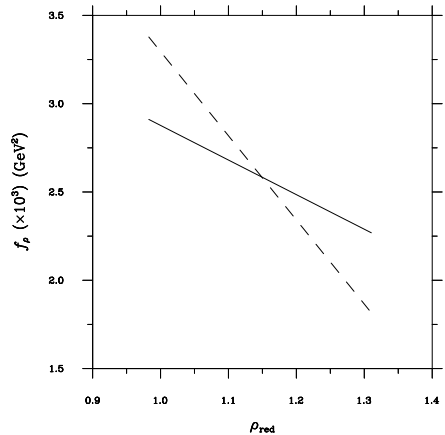,height=2.5in,width=2.5in}}
\vspace{10pt}
\caption{Dependence of the $\rho$ spectral parameter, $f_\rho$, on the
reduced VSA-violation parameter, $\rho_{red}$, for the two weight
families, $w_s$ and $w_d$.  The solid line corresponds to $w_s$,
the dashed line to $w_d$.}
\label{fig1}
\end{figure}

Having extracted $f_V$, and knowing the EM decay constants,
$F^{EM}_V$, one may then separately determine
$F^3_V$ and $F^8_V$ for $V=\rho ,\omega ,\phi$.  
To understand the impact of the IB decay constants on the extraction
of the $33$ and $88$ spectral functions, it is convenient
to quote the results in terms of the squared ratios, $r_V$, shown below.
In each case, the numerator in the ratio of decay constants
represents the IC decay constant
of the meson in question (whose square is relevant to the contribution
to either the $33$ or $88$ spectral function), while the denominator
represents the nominal value of the numerator, obtained from EM
data neglecting the presence of the 
``wrong-isospin'' IB contributions discussed above.  
The deviations of the $r_V$ from $1$ reflect the presence of
the IB $F^8_\rho$, $F^3_\omega$ and $F^3_\phi$ contributions
to $F^{EM}_{\rho ,\omega ,\phi}$.
The ratios have been defined in such a way that the true
resonance contributions to the $33$ (or $88$) spectral functions 
are obtained by multiplying 
the nominal results, obtained in the usual analysis, by $r_V$.
The results of the analysis outlined above are then
\begin{eqnarray}
r_\rho &=&\left[ \frac{F^{3}_\rho}{F^{EM}_\rho}\right]^2= 0.982\pm 0.0021
\nonumber \\
r_\omega &=&\left[ \frac{F^{8}_\omega}{\sqrt{3}\, F^{EM}_\omega}\right]^2=
1.154\pm 0.017 \nonumber \\
r_\phi &=&\left[ \frac{F^{8}_\phi}{\sqrt{3}\, F^{EM}_\phi}\right]^2=
1.009\pm 0.001 \ ,
\label{corrections}
\end{eqnarray}
where the errors are dominated by the uncertainty in the quark
mass ratio, $r$, and hence are strongly correlated.

We note the following features of the above results:
\begin{itemize}
\item The effect on the $\rho$ contribution to the $33$ spectral function
is small, and hence has no impact on CVC tests, given the current
experimental errors on the EM cross-sections.
\item As expected from the argument above, the effect is much larger 
for the $\omega$ contribution to the $88$ spectral function.  Our
result corresponds to a $\sim 7.5\%$ isovector contribution to the physical
$\omega$ decay constant.
\item The ratio of $\rho$ and $\omega$ corrections is $8.6\simeq 9$,
suggesting dominance by $\rho$-$\omega$ mixing.
\end{itemize}

The impact of these corrections on the $33$-$88$
sum rules discussed above is as follows.
\begin{itemize}
\item Using $\tau$ decay data for the isovector part, and the above
corrections for the EM isoscalar data, the Narison $m_s$ sum rule implies
\begin{equation}
m_s(1\ {\rm GeV}^2)=154(\pm\sim 50) \ {\rm MeV}\ ,
\end{equation}
compatible with value $157\pm 16\pm 15\pm 16$ obtained in a recent
analysis based on flavor breaking in hadronic
$\tau$ decays~\cite{jkkm00}, and to be compared to the central
value $197$ MeV obtained neglecting the presence of IB 
corrections\cite{narisonms}.
\item The correction to the inverse chiral $33-88$ sum rule extraction of the
$6^{th}$ order ChPT low-energy constant $Q_V$ is also large.
The FESR solution above implies, via an inverse moment $38$ sum rule, the
value~\cite{kmcewv3v8}
\begin{equation}
Q_V(m_\rho^2)=(3.3\pm 0.4)\times 10^{-5}\ ,
\label{fesrqv}
\end{equation}
to be compared to the value from a 
recent analysis~\cite{jksd99} of the vector current
$\Pi^{ud}$-$\Pi^{us}$ difference,
employing $\tau$ decay spectral data (which does not need IB corrections) 
\begin{equation}
Q_V(m_\rho^2)=(2.8\pm 1.3)\times 10^{-5}\ .
\label{fesrnewqv}
\end{equation}
The agreement of the results of Eq.~(\ref{fesrqv}) with the independent
determination of Eq.~(\ref{fesrnewqv}) provides further strong support
for the values of the IB vector meson decay constants obtained in the
present analysis.
\end{itemize}

\end{document}